\documentstyle[12pt]{article}

\catcode`\@=11
\font\twelvemsb=msbm12
\newfam\msbfam
\textfont\msbfam=\twelvemsb
\edef\msbfam@{\hexnumber@\msbfam}
\def\Bbb#1{{\fam\msbfam\relax#1}}

\newcommand{\M}{{\cal M}}
\newcommand{\V}{{\cal V}}
\newcommand\T[1]{(S_{#1})}
\newtheorem{lemma}{Lemma}
\newtheorem{theorem}{Theorem}
\newtheorem{definition}{Definition}

\newcommand\bm[1]{\mbox{\boldmath$#1$}}
\newcommand\tr[1]{\mbox{Tr$#1$}}
\newcommand{\n}{\bm{\ast}}
\begin{document}
\title{3+1 description of silent universes: a uniqueness result
for the Petrov type I vacuum case.}
\author{Marc Mars\thanks{Also at Laboratori de
F\'{\i}sica Matem\`atica, Societat
Catalana de F\'{\i}sica, IEC, Barcelona, Spain.}\\ 
Institute of Theoretical Physics, University of Vienna \\
Boltzmanngasse 5, A-1090 Vienna, Austria.}
\maketitle
\begin{abstract}
Silent universes are studied using a ``3+1'' decomposition of the
field equations in order to make progress in proving a 
recent conjecture that the only silent universes of Petrov type
I are spatially homogeneous Bianchi I models. The infinite set
of constraints are written in a geometrically clear
form as an infinite set of Codacci tensors on the initial
hypersurface. In particular, we show
that the initial data set for silent universes is ``non-contorted'' and
therefore (Beig and Szabados \cite{Beig})
isometrically embeddable in a conformally flat spacetime. 
We prove, by making use of algebraic computing programs,
that the conjecture holds in the simpler case when the spacetime is
vacuum. This result points to confirming the
validity of the conjecture in the general case.
Moreover, it provides an invariant
characterization of the Kasner metric directly in terms
of the Weyl tensor. A physical interpretation of this uniqueness result
is briefly  discussed.
\end{abstract}
PACS Numbers: 04.20.Ex, 02.40
\vspace{1cm}
\newpage

\section{Introduction}

In the last few years there has been intense activity in the study
of silent universes, which are dust
spacetimes such that the fluid velocity vector is irrotational and the
magnetic part of the Weyl tensor with respect to this vector vanishes. 
The concept was first put forward by 
S.Matarrese, O.Pantano and D.S\'aez \cite{MPS} in an
astrophysical and cosmological context in order to describe
structure formation in the universe.
Moreover, silent universes were interesting also from an
exact solutions point of view because
this class contains the Szekeres family of spacetimes \cite{Sze}, which has
been extensively analyzed in the literature (see \cite{Kras} for a
comprehensive review). The Szekeres class 
exhausts the silent universes of Petrov type D. Since
silent universes are necessarily of Petrov type 0, D or I
(Barnes and Rowlingson \cite{BR}) and the conformally
flat case is well understood, the interesting new physics was in the
Petrov type I subclass
and a number of relevant new results were expected.
This fact triggered active research on silent universes
\cite{sil1}-\cite{siln}.
In particular, it was soon realized that the Einstein 
equations and the
Bianchi identities, when written in a suitable orthonormal tetrad,
decouple into a set of ordinary differential equations (which describe the
time evolution of the spacetime) and a set of constraint equations (which
only involve derivatives along spatial directions). Most of the studies
focussed in analyzing the
evolution equations, mainly  by making use of a dynamical system formulation
of the Einstein field equations (see \cite{Wain} for a detailed account
of the method). The existence problem of silent universes of
Petrov type I was apparently settled in \cite{Lesame} where it was claimed
that solving the constraint equations on an initial hypersurface
is necessary and
sufficient to obtain a silent universe
(or, equivalently, that the time evolution
of the constraints is automatically satisfied if the constraints hold
initially). Unfortunately, this claim turned out to be untrue
\cite{Mars}. So, the problem of how large
the class of Petrov type I silent universes is had to be reanalyzed. In two
independent works \cite{Carlos}-\cite{Henk} this issue was addressed. The
analysis of the time-evolution of the constraints was performed in 
\cite{Carlos} using a coordinate approach and in \cite{Henk} using
a tetrad approach. In both cases it was found that the 
successive time derivatives of the constraints become larger and larger
expressions which are algebraically independent from each other. This led
these authors to conjecture that silent universes of Petrov type I are
extremely scarce and that the whole class reduces to 
some Bianchi models. Proving this conjecture turns out to be 
a very
difficult problem because the successive constraints become huge
just after a few time derivatives
and they become unmanageable even for algebraic
computing programs. Our aim in this paper is twofold. First, we perform
a study of silent universes using the ADM splitting \cite{ADM}
with respect to the hypersurface $\Omega$ orthogonal to the fluid velocity.
The main advantage of this method 
is that the full set of constraints can be written in a 
geometrically neat
way as an infinite sequence of symmetric tensors
satisfying the Codacci equation (i.e. a sequence of so-called
Codacci tensors). Moreover, we show that the
initial data set $(\Omega, h_{ab}, K_{ab})$ satisfies the so-called
{\it non-contorted} condition. R.Beig and L.B.Szabados \cite{Beig} 
have recently proven that, locally, non-contorted initial data sets
can be isometrically embedded in a conformally 
flat spacetime. Thus, each hypersurface orthogonal to the
fluid velocity vector in a silent universe can be locally viewed as a
hypersurface in a conformally flat spacetime. We have not been able to
exploit this fact fully yet, but we believe that an appropriate use of this
result could prove essential to settle the question on the non-existence
of spatially inhomogeneous silent universes of Petrov
type I. However, this question is not analyzed here any further. Rather, we
concentrate in the second objective of the paper. The great
complexity of analyzing the
set of constraints in the general case simplifies 
in the vacuum case (essentially because there are five unknown functions
instead of six). This allows us to prove,
by making use of algebraic computing, that
a vacuum silent universe
of Petrov type I must be locally isometric to the Kasner spacetime \cite{Kas}
(which is a one-parametric family of  spatially homogeneous
Bianchi I vacuum spacetimes) thus showing that the conjecture holds in vacuum.
Although the concept of silent universe was originally put forward
for dust spacetimes, the definition makes obviously sense for vacuum
spacetimes as well. The only difference is that in vacuum there
is no privileged timelike congruence, unlike in the dust case, and the
definition must be modified accordingly (see Definition \ref{silent}).
Since Petrov type D vacuum silent universes 
are just the vacuum subclass of the Szekeres family,
and vacuum conformally flat silent universes are, of course,
locally Minkowski, our result provides a complete
classification of vacuum silent universes. This result is interesting
at least in three respects.
First, it  suggests the validity of the conjecture
in the general case.
Furthermore, it provides an invariant characterization of the Kasner
spacetimes without involving isometries (the Kasner family is
well-known to be the most general spatially homogeneous Bianchi type I
vacuum metric).
Finally, this result can also be interpreted physically as follows.
The uniqueness result we prove states that non-trivial
vacuum silent spacetimes must be of Petrov type D (or 0) (we consider Kasner
as trivial due to its high degree of symmetry).
Gravitational fields of Petrov type D have often been considered analogous
to Coulombian electromagnetic fields in flat space. This analogy is,
obviously, loose because the Petrov type of a spacetime 
is a local property, while characterizing an electromagnetic
field as Coulombian involves conditions on the decay rate near
spatial infinity, which is a global
property. Nevertheless, accepting this
analogy and considering Petrov type D gravitational
fields as ``locally Coulombian'',  our uniqueness 
result could be interpreted physically as establishing that
vacuum gravitational fields which are purely electric with
respect to an inertial observer must be of Coulombian type (except
for the Kasner metric, which is a special case). A similar result also holds
in electromagnetism in flat space, where it is true 
that electromagnetic fields
which are purely electric with respect to an inertial observer must be
Coulombian. However, this analogy should be taken cautiously because
the terms ``inertial'' and ``Coulombian'' have a local meaning 
in gravity (corresponding to `` geodesic irrotational 
observer'' and ``Petrov type D'', respectively) while
they are global in flat space (meaning ``orthogonal to hyperplanes''
and ``with decay rate as $r^{-2}$ at infinity''). Nevertheless, we
believe that this
analogy provides some physical explanation for the uniqueness result
we have obtained.

The paper is organized as follows. In section 2 we describe briefly the
ADM formalism for an arbitrary energy-momentum tensor. This fixes
our notation and conventions. In section 3 we adapt these
equations to the silent universe case. In particular, we show that
the evolution equations can be written as a system of ordinary
differential equations when suitable variables are chosen. Then,
we concentrate in the study of the constraint equations. We prove that
the whole set of constraints take the form of an infinite sequence
of tensors $\T{n}^{a}_{b}$ which satisfy the so-called Codacci equation.
The tensors $\T{n}^{a}_{b}$ depend algebraically on the second fundamental
form $K_{ab}$ and the Ricci tensor of $h_{ab}$. Then, we show that
the Codacci equation for $\T{0}_{ab}$ and $\T{1}_{ab}$ imply that the 
initial data set is non-contorted 
and quote the result by Beig and Szabados mentioned above.
In Section 4 we study the
Codacci equation by rewriting it as a Pfaffian system
for one-forms. This allows us to study the integrability conditions
of this equation easily and show that 
$\T{n}_{ab}$ are symmetric and commute with each other.
Using these results, the evolution
equations are rewritten in terms of the eigenvalues of $\T{0}^{a}_{b}$ and
$\T{1}^{a}_{b}$. In section 5 we restrict ourselves to the vacuum case
and we prove our main theorem, namely that an
arbitrary vacuum silent universe of Petrov type I is locally isometric
to Kasner. The computations necessary to show this result have been
performed using Reduce. However, the proof is not straightforward and
needs some discussion. Hence we have included the source code 
of the program, as well as the necessary explanations in Appendix A.
Reading this Appendix in detail requires some knowledge of Reduce but
we believe that  the general idea of the proof can be understood also
by the non-expert just by following the comments we have included.

\section{Preliminaries and basic results}

In this section we review briefly the ADM formalism \cite{ADM} 
and write down its basic equations. This will fix our notation
and conventions.

The study of silent universes
was motivated, among other reasons, by the fact that the Einstein
field equations split into a set of {\it ordinary}
differential equations (the evolution equations)
and a set of constraint equations. In order to see why this
happens, let us start by writing down the ``3+1'' equations for
an arbitrary energy-momentum tensor. Consider
a smooth irrotational timelike congruence of curves on a spacetime
$(\M,g)$\footnote{For simplicity we assume the spacetime to be
$C^{\infty}$ although lower differentiability would suffice. The
signature of the metric $g$ is $(-1,1,1,1)$, the Levi-Civita covariant
derivative is denoted by $\nabla$ and our
sign conventions of the Riemann and Ricci tensor
follow \cite{KSMH}} with unit tangent vector $\vec{u}$. Then, 
there exists an embedded hypersurface $\Omega$ which is orthogonal
to $\vec{u}$, i.e. $\Omega$ is a three-dimensional
manifold and $\varphi_0 : \Omega \,\rightarrow \, \M$ 
is an embedding such that
$\varphi_0^{\star} (\bm{u}) = 0$ ($\bm{u}$ is obtained by lowering the index to
$\vec{u}$ and the $\star$ denotes the pullback, as usual).
Let us choose an
arbitrary point $p \in \Omega$ and an open, connected
neighborhood $U_p \subset \Omega$ of $p$ with compact closure. 
Let $t$ denote the proper-time parameter
of the congruence satisfying $t=0$ on $\varphi_0(U_p)$.
Then, there exists a positive real number $t_0$ such that the integral
curves of $\vec{u}$ crossing $\varphi_0(U_p)$ exist for 
$t \in I_0=(-t_0,t_0) \subset \Bbb{R}$ 
and do not intersect $\varphi_0(U_p)$ again (also for $|t|<t_0$). 
Let $\Phi_t$ denote the one-parameter
local group of transformations generated by $\vec{u}$. The map 
$\varphi_t = \Phi_t \circ \varphi_0|_{U_p}$ is an embedding 
$\varphi_t \, : U_p \rightarrow \M$ and 
$\varphi_t(U_p)$ is a spacelike hypersurface in $\M$. 
The set $\M_p \equiv \bigcup_{|t|<t_0} \varphi_{t} (U_p)$ is
diffeomorphic to $U_p \times I_0$ and ($\M_p,g|_{\M_p}$) is 
globally hyperbolic.
Given an r-index covariant tensor field $T$ on $\M_p$ we define
a one-parameter family of covariant tensor fields on $U_p$ by
$ T_{j_1 \cdots j_r} (t) \equiv \varphi_t^{\star} \left (T
\right)_{j_1 \cdots j_r}$, $t \in I_0$ (tensors on $U_p$ carry Latin
indices and tensors on $\M$ carry Greek indices).
The definition of Lie derivative yields
$\varphi_t^{\star} \left ( {\cal L}_{\vec{u}} T
\right )_{j_1 \cdots j_r} = \partial_t T_{j_1 \cdots j_r}(t)$.  
Define as usual the acceleration and the deformation tensor by
$a_{\mu} = u^{\nu} \nabla_{\nu} u_{\mu}$ and
$\Sigma_{\mu\nu} = h_{\mu}^{\,\,\alpha} h_{\nu}^{\,\,\beta} \nabla_{\alpha}
u_\beta$, where $h_{\mu}^{\,\,\nu} = \delta_{\mu}^{\,\,\nu} + u_{\mu} u^{\nu}$
is the projector. $U_p$ can be endowed with a family of
symmetric tensors $h_{ab}(t)$ and $K_{ab}(t)$, $t \in I_0$ by
\begin{eqnarray*}
h_{ab} (t) = \varphi_t^{\star} \left (g \right )_{ab},
\hspace{1cm}
K_{ab} (t) = \varphi_t^{\star} \left (\Sigma \right )_{ab}.
\end{eqnarray*}
For fixed $t$, $h_{ab}(t)$ is a positive definite metric on $U_p$.
We denote the corresponding Levi-Civita covariant derivative
by $D^t$ and the Riemann and Ricci tensors by $R_{abcd}(t)$ and $R_{ab}(t)$.
For objects at $t=0$ we will drop the argument, i.e.
$h_{ab}(0)$, $K_{ab}(0)$ and $D^0$ will be written simply as $h_{ab}$,
$K_{ab}$ and $D$ respectively (and similarly for other tensors
on $U_p$).
The Gauss and Codacci identities relating the Riemann tensors of 
$(\M_p, g|_{\M_p} )$ and $(U_p, h_{ab}(t) )$ are \cite{KN}
\begin{eqnarray}
\varphi_t^{\star} ({\cal R})_{abcd} & = & R_{abcd}(t) 
+ K_{ac}(t) K_{bd}(t) - K_{ad}(t) K_{bc}(t), \label{Gauss}\\
\varphi_t^{\star} \left ({\cal R}(\vec{u}) \right )_{abc} & = &
D^{t}_{c} K_{ab} (t) - D^{t}_{b} K_{ac} (t), \label{Codacci}
\end{eqnarray}
where ${\cal R}$ is the Riemann tensor
of $(\M_p,g|_{\M_p})$ with all the
indices lowered, and ${\cal R} (\vec{u})$ is the tensor 
${\cal R} (\vec{u})_{\alpha\beta\gamma} = 
u^{\delta} {\cal R}_{\delta\alpha\beta\gamma}$.
Since the Ricci and the Riemann tensors are equivalent 
in three dimensions, we can contract (\ref{Gauss})
with $h^{ac}(t)$
without loss of information. The result is best written
by introducing the electric and magnetic parts of
the Weyl tensor in $(\M_p,g|_{\M_p})$ as
$E_{\alpha\beta} = u^{\mu} u^{\nu} C_{\mu\alpha\nu\beta}$,
$H_{\alpha\beta} = \frac{1}{2} u^{\mu} u^{\nu}
\eta_{\mu\alpha\tau\sigma} C^{\tau\sigma}_{\;\;\;\;\;\nu\beta}$,
($\eta_{\alpha\beta\gamma\delta}$ is the volume form of the spacetime).
Using now the Einstein field equations for an arbitrary energy-momentum
tensor, we can decompose the Ricci tensor of $(\M_p, g|_{\M_p})$ as
\begin{eqnarray*}
{\cal R}_{\alpha\beta} = 
\Pi_{\alpha\beta} + \left (\rho + p \right ) u_{\alpha}
u_{\beta} + \frac{1}{2} \left (\rho - p \right ) g_{\alpha\beta}
- u_{\alpha} q_{\beta} - u_{\beta} q_{\alpha}.
\end{eqnarray*}
where $\Pi_{\alpha\beta}$ is symmetric and trace-free and
$\Pi_{\alpha\beta}$, $q_{\alpha}$ are orthogonal to $\vec{u}$.
Let us define $E_{ab}(t)$, $H_{ab}(t)$,
$\Pi_{ab}(t)$, $q_a(t)$, $a_b(t)$, $\rho(t)$ and $p(t)$ as
the pullbacks with respect to $\varphi_{t}$ of $E_{\alpha\beta}$,
$H_{\alpha\beta}$,
$\Pi_{\alpha\beta}$, $q_{\alpha}$, $a_{\beta}$, $\rho$ and $p$ respectively.
The contracted Gauss identity reads
\begin{eqnarray}
E_{ab} (t) + \frac{1}{2} \Pi_{ab} (t) + \frac{2}{3} \rho (t)
h_{ab} (t) = 
R_{ab}(t) + K(t) K_{ab}(t) - K_{ac}(t) K^{c}_{\;b} (t) \equiv 
\label{Gausscont} \\
\hspace{5cm} \equiv S_{ab} (t) + S(t) h_{ab} (t), \nonumber
\end{eqnarray}
where $K(t) \equiv h^{ab}(t) K_{ab}(t)$, $S(t) \equiv h^{ab}(t)S_{ab}(t)$,
Latin indices are raised with $h_{ab}(t)$ and
the symmetric tensor $S_{ab}(t)$ is defined through the second identity.
The Codacci identity (\ref{Codacci}) can be written as
\begin{eqnarray}
\eta^t_{\;dbc} \, H^{d}_{\;a}(t) +
\frac{1}{2} \left ( \frac{}{} h_{ac}(t) \, q_b(t) - h_{ab}(t) \, q_c (t)
\right ) = D^t_{c} K_{ba}(t) - D^t_{b} K_{ca}(t) \equiv
Z_{cba}(t). \label{Codcont}
\end{eqnarray}
where $Z_{cba}(t)$ is defined by the second identity and 
$\eta_{abc}^t$ stands for the volume form of $(U_p,h_{ab}(t))$.
Applying $\varphi^{\star}_{t}$ to the trivial identity
${\cal L}_{\vec{u}} \, g_{\alpha\beta} = \nabla_{\alpha} u_{\beta} +
\nabla_{\beta} u_{\alpha}$ yields
\begin{eqnarray}
\frac{\partial h_{ab} (t)}{\partial t} = 2 K_{ab} (t). \label{dth}
\end{eqnarray}
Finally, applying $\varphi^{\star}_{t}$ to
${\cal L}_{\vec{u}} \Sigma_{\mu\nu} = \Sigma_{\mu\beta}
\Sigma^{\beta}_{\,\,\nu}
+ a_{\mu} a_{\nu} + h_{\mu}^{\,\,\alpha} h_{\nu}^{\,\,\beta}
\nabla_{\alpha} a_{\beta} - h_{\mu}^{\,\,\alpha} u^{\beta}
h_{\nu}^{\,\,\gamma} u^{\delta} R_{\alpha\beta\gamma\delta}$ (which
is a well-known, and in any case easily verifiable, identity) we 
get, after raising one index and using (\ref{Gausscont}),
\begin{eqnarray}
\frac{\partial K^{b}_{\;c} (t)}{\partial t} 
 =  - K^{b}_{\;d}(t) K^{d}_{\;c} (t) +
a^{b}(t) a_{c}(t)+ D^t_{c} a^b(t) - S^{b}_{\;c}(t) +
\Pi^{b}_{\;c}(t) - \frac{p(t)}{2} \delta^{b}_{\;c}.
\label{dtk}
\end{eqnarray}
The geometrical identities (\ref{Gausscont}),(\ref{Codcont}) are the
so-called constraint equations and (\ref{dth})-(\ref{dtk})
are the evolution equations.

\section{ADM formalism for silent universes}

The evolution equations (\ref{dth})-(\ref{dtk}) are, in general,
partial differential equations. The main property  of
silent universes is that, using an appropriate set of variables,
the evolution equations become 
{\it ordinary} differential equations. To show this,
we first write down the evolution equation for $R^{a}_{\;b}(t)$.
Let ${\Gamma}^{a}_{bc}(t)$ denote the Christoffel symbols
of $h_{ab}(t)$. An easy consequence of (\ref{dth}) is 
$\partial_t \Gamma^{a}_{bc}(t)= D^{t}_b K_{\;c}^{a} (t) 
+ Z_{c\;\;\;b}^{\;\;a}(t)$. The identity
$\partial_t R^{a}_{\;bcd} (t)
= D_{c}^{t} \left [ \partial_t \Gamma^{a}_{bd}(t) \right ] -
D_{d}^{t} \left [ \partial_t \Gamma^{a}_{bc} (t) \right ]$ becomes, after
using the Ricci identity,
\begin{eqnarray}
\frac{\partial R^{a}_{\;b} (t)}{\partial t} = \left .
\left [ D_{c} Z^{ac}_{\;\;\;\;b} - D_{b} Z^{a c}_{\;\;\;\;c}  +
R^{a}_{\;c} K^{c}_{\;b} - K R^{a}_{\;b}  -
\frac{1}{2} R  K^{a}_{\;b} + \delta^{a}_{\;b}
\left (\frac{1}{2} R K - \tr{(RK)} \right ) \right ] \right |_t
\label{dtR}
\end{eqnarray}
where $\tr{(RK)}(t) = R^{a}_{\;b}(t) K^{b}_{\;a} (t)$
and ``$\left . \right |_t$'' means that all the objects enclosed
(including the covariant derivative) are to
be taken at the value $t$. The crucial
fact in (\ref{dtR}) is that spatial derivatives act only on 
the tensor $Z_{abc}(t)$. Hence, by imposing
conditions on this object it is possible to obtain
a system of ordinary differential equations for 
$K^{a}_{\;b}(t)$ and $R^{a}_{b}(t)$. The simplest possibility
is to set $Z_{abc}(t) \equiv 0$, which, from equation (\ref{Codcont}),
is equivalent to  $q_a(t) \equiv 0$ and $H_{ab}(t) \equiv 0$. 
Moreover, (\ref{dtk}) shows that some evolution equations for
$a_c(t)$, $p(t)$ and $\Pi_{ab}(t)$ are needed in order to
obtain a closed system of differential equations.
Again, the simplest possibility is
that they vanish identically. This motivates the standard definition
of silent universe.
\begin{definition}
\label{silent}
A spacetime $(\M,g)$ is called a silent universe if it admits a
unit timelike vector field $\vec{u}$ with is irrotational and geodesic,
the magnetic part of the Weyl tensor with respect to $\vec{u}$
vanishes and the Ricci tensor takes the form
${\cal R}_{\alpha\beta}= \rho \left (u_{\alpha}
u_{\beta} + \frac{1}{2} g_{\alpha\beta} \right )$, where $\rho$ is an
arbitrary smooth function (possibly zero).
\end{definition}
The condition that $\vec{u}$ is geodesic is superfluous in the
dust case ($\rho \neq 0$) because of the contracted Bianchi
identities, but
it must be required additionally in the vacuum case.
The term ``silent'' stems precisely from the fact that the evolution equations
become ordinary differential equations and hence, no influence from
neighbouring points (apart from the one encoded in the initial
data) arises during evolution. 

From now on, the spacetime $(\M,g)$ will denote a silent
universe. Then, the Codacci identity becomes 
\begin{eqnarray}
Z^{\;\;\;c}_{ab}(t)=
D^{t}_{a} K^{c}_{\;b} (t)- D^{t}_{b} K^{c}_{\;a} (t)= 0
\label{const}
\end{eqnarray}
and the evolution equation for $K^{a}_{\;b}(t)$ reads
\begin{eqnarray}
\label{dtk2}
\frac{\partial K^{b}_{\;c} (t)}{\partial t} 
 = - \left .
\left (  K^{b}_{\;d} K^{d}_{\;c}  + S^{b}_{\;c} \right ) \right |_t .
\end{eqnarray}
Regarding the remaining evolution equation, it turns
out to be more convenient to use $S^{a}_{\;b}(t)$ 
instead of $R^{a}_{\;b}(t)$. A straightforward, if somewhat long,
calculation using (\ref{dtR}) and (\ref{dtk2}) yields
\begin{eqnarray}
\frac{\partial S^{a}_{\,b}(t)}{\partial t}
 = \left .  \left [ 2 S^{a}_{\,c} K^{c}_{\,b} + 
 K^{a}_{\,c} S^{c}_{\,b} - 2 S K^{a}_{\, b} -
2 K S^{a}_{\, b} + \delta^{a}_{\,b} 
 \left ( S K  - \tr{\left (S K \right )} \right ) \frac{}{} 
\right ] \right |_t,
\label{dtS}
\end{eqnarray} 
where $\mbox{Tr}(S K)(t) = S^{a}_{\,b}(t) K^{b}_{\,a}(t)$.
In order to derive this equation the following algebraic identity
(which is valid for
any $3 \times 3$ matrix $A$) has been used 
\begin{eqnarray*}
A^3 = \mbox{tr} (A)  A^2 + \frac{ \mbox{Tr} (A^2) -
\mbox{Tr}^2 (A)}{2} A + \mbox{I}_3 \left [
\frac{1}{3} \mbox{Tr} (A^3 ) - \frac{1}{2} \mbox{Tr} (A)
\mbox{Tr} (A^2) + \frac{1}{6} \mbox{Tr}^3 (A) \right ].
\end{eqnarray*}
Thus, the evolution equations for silent universes are indeed very
simple (the definition was designed for this purpose). In return,
the set of constraints become highly non-trivial. Indeed, the set of
constraints $Z_{abc}(t)=0$ provide, after time differentiation,
new constraints which must be satisfied identically. Let
us describe this in detail. For the initial data problem, a triple 
$(U_p, h_{ab}, K_{ab})$ satisfying the Codacci constraint
$D_{\left [a \right .} K^{b}_{\left . \,\, c \right ]} = 0$ must be given.
Then, the tensor $S_{ab}$ is defined via $S_{ab} + S h_{ab} \equiv
R_{ab} + K K_{ab} - K_{ac} K^{c}_{\;b} $.
Next, $K^{a}_{\;b}(t)$ and $S^{a}_{\;b}(t)$ are obtained as the
unique solutions of the ordinary
differential system (\ref{dtk2})-(\ref{dtS}) 
with initial data $K^{a}_{\;b}(0)=K^{a}_{\;b}$ and 
$S^{a}_{\;b} (0) = S^{a}_{\;b}$. Afterwards
$h_{ab}(t)$ can be found as the unique solution of $(\ref{dth})$ (the
right hand-side is already known) satisfying
$h_{ab}=h_{ab}(0)$. It is easy to show that the tensor
$S^{a}_{\;b}(t)$ constructed  a posteriori from
$h_{ab}(t)$ and $K_{ab}(t)$ coincides with the
solution of (\ref{dtS}) we started from. It remains the check as to
whether  the Codacci equation for $K^{a}_{\;b}(t)$ is fulfilled for all $t$.
A necessary condition is that all the time derivatives
of $Z_{abc}(t)$ vanish at $t=0$ (whether this is also sufficient would
require extra, non-trivial work). In \cite{Carlos} and
\cite{Henk} the time derivatives were discussed in a coordinate and tetrad
setting respectively, and they were found to become increasingly large and
intractable (the explicit expressions for the first few
constraints appear in \cite{CarT}) but almost nothing about
the geometric structure of the successive constraints was revealed.
The ADM formalism
turns out to be more useful for this purpose. To that aim, let us find an
expression
for the commutation between  $\partial_t$ and $D^{t}$. Let
$N^a_{\;b}(t)$ be any one-parameter family of tensors on $U_p$. 
A simple calculation yields
\begin{eqnarray}
\label{commun}
\frac{\partial}{\partial t} \left (D_a^t N^b_{\;c}(t) \right ) =
\left . \left [  D_a \left (\frac{\partial}{\partial t}
 N^b_{\;c}  \right ) +
N^{d}_{\;c}  \, D_a K^{b}_{\;d}  
- N^b_{\;d}  \, D_{a} K^{d}_{\;c} \, \right ]\right |_t. 
\end{eqnarray}
\begin{lemma}
\label{recur}
Let $(\M,g)$ be a silent universe and $(U_p,h_{ab}(t),K_{ab}(t))$
constructed as in Sect. 2. Let $A^{b}_{\,\,c} (t)$ be a family of tensors
on $U_p$ satisfying $D^{t}_{\left [a \right .} A^{b}_{\left . 
\,\, c \right ]}(t) = 0$  $\forall t$. Then the family of tensors 
\begin{eqnarray*}
B^{b}_{\,\,c} (t) =  \partial_{t}  A^{b}_{\,\,c} (t)+
K^{b}_{\,\,d}(t) A^{d}_{\,\,c}(t)
\end{eqnarray*}
also satisfies
$D^{t}_{\left [a \right .} B^{b}_{\left . \,\, c \right ]}(t) 
= 0$ $\forall t$.
\end{lemma}
{\it Proof}.
Taking the time derivative of $D^{t}_{\left [a \right .} A^{b}_{\left . 
\,\, c \right ]}(t) = 0$  and using the commutation formula (\ref{commun})
we find
\begin{eqnarray*}
0 = \left .\frac{\partial}{\partial t} \left (
D_{\left [a \right . } A^{b}_{\,\,\left . c \right ]} \right ) \right |_t
= \left . \left [
D_{\left [a \right . } \left (\frac{\partial}{\partial t}
A^{b}_{\,\,\left . c \right ]} \right ) +
A^{d}_{\left [ \; c \right .} 
D_{\left . a \right ]} \left ( K^{b}_{\; d } \right ) \right ] \right |_t =
\left .  D_{\left [ a \right . } \left (
A^{d}_{\; \left .  c \right ]} K^{b}_{\; d} +
\frac{\partial}{\partial t} A^{b}_{\;\left . c \right ]} \right ) \right |_t
\end{eqnarray*}
where $D^{t}_{\left [a \right .} K^{b}_{\left . 
\,\, c \right ]}(t) = 0$ was used in the second  equality and
$D^{t}_{\left [a \right .} A^{b}_{\left . 
\,\, c \right ]}(t) = 0$ in the third one. $\hfill \Box$

This lemma, together with the constraint equations (\ref{const}) suggests
defining the following sequence of one-parameter families of
tensors on $U_p$
\begin{eqnarray}
{\T{0}}^{a}_{\;b} (t) = -K^{a}_{\;b} (t),
\hspace{1cm}
{\T{n}}^{a}_{\;b} (t) = 
\frac{\partial}{\partial t} \left [
\left . {\T{n-1}}^{a}_{\;b}(t) \right . \right ]
+ K^{b}_{\,\,d}(t) {\T{n-1}}^{d}_{\,\,c}(t),
\quad n \in \Bbb{N}.
\label{defCod}
\end{eqnarray}
Lemma \ref{recur} shows that ${\T{n}}^{a}_{\;b}(t)$ satisfies
$D^{t}_{\left [a\right .} {\T{n}}^{c}_{\left .\;b
\right ]} (t) = 0$, $\forall n \in \Bbb{N}\cup \{0\}, \forall t$.
Hence, the successive time derivatives of  $Z_{abc}(t)=0$
at $t=0$ can be written as 
 $D_{\left [a\right .} {\T{n}}^{c}_{\left .\;b \right ]} = 0$,
$\forall n \in \Bbb{N}\cup \{0\}$ and these are the full set of constraints
that the initial data $(U_p,h_{ab},K_{ab})$ must satisfy.

Notice that the evolution equation (\ref{dtk2}) implies 
${\T{1} }^{a}_{\;b}(t) = S^{a}_{\;b}(t)$. Furthermore, if follows from
(\ref{dtk2}) and (\ref{dtS}) that each
${\T{n}}^{a}_{\;b}(t)$ depends algebraically on $K^{a}_{\;b}(t)$ and
$S^{a}_{\;b}(t)$. Moreover, it follows easily that 
this dependence is polynomial and that if we
decide that $K^{a}_{\;b}(t)$ carries a degree equal to one and
$S^{a}_{\;b}(t)$ carries a degree equal to two, then 
${\T{n}}^a_{\;b}(t)$ is homogeneous of degree $n+1$ (i.e. it
consists of a sum of terms of degree $n+1$).
Unfortunately, obtaining an explicit general formula for 
${\T{n}}^a_{\;b}(t)$ in terms of $K^{a}_{\;b}(t)$ and
$S^{a}_{\;b}(t)$ seems to be a difficult task. The first two
constraint equations read simply
\begin{eqnarray*}
D_{\left [a\right .} {K}^{c}_{\left .\;b \right ]}  = 0,
\hspace{1cm}
D_{\left [a\right .} {S}^{c}_{\left .\;b \right ]}  = 0.
\end{eqnarray*}
In particular, this shows that the initial data set $(U_p,
h_{ab},K_{ab})$ is non-contorted (see \cite{Beig} for a definition).
In that paper, the authors prove that any non-contorted initial data
set can, locally, be isometrically embedded in a conformally flat
spacetime. Hence, each hypersurface orthogonal to the
fluid flow in
a silent universe can be locally embedded in a conformally flat
spacetime. This already gives substantial geometric information about
the initial data set (and only the first two constraints have been used!).
We believe that this fact can provide the key for proving the conjecture
that silent universes of Petrov type I must be spatially homogeneous 
(in terms of the initial data set, being spatially homogeneous
is equivalent to $(U_p,h_{ab})$ being flat and $K_{ab}$ being covariantly
constant). However we have not been able to exploit this fact fully 
yet and the question remains under investigation.

Summarizing, in this section we have obtained the evolution
equations for the silent
universes as ordinary differential equations in terms of the
tensors $K^{a}_{\;b}(t)$ and $S^{a}_{\;b}(t)$ and we have written
the full set of constraints  in the form
$D^{t}_{\left [a \right .} {\T{n} }^{b}_{\left . \,\, c \right ]}(t) 
= 0$ $\forall t$,
for a collection of tensors $\T{n} {}^a_{\;b}(t)$ which depend
algebraically on $K^{a}_{\;b}(t)$ and
$S^{a}_{\;b}(t)$. Since all these tensors satisfy the same equation,
the next section is devoted to study it in some detail.

\section{Consequences of the Codacci equation}

Let us start with a standard definition 
\begin{definition}
Let $(\V,\gamma)$ be a (pseudo-)Riemannian manifold
and $D$ the Levi-Civita covariant derivative. A symmetric
tensor field $Q_{ab}$ is called a Codacci tensor iff it satisfies the
Codacci equation
\begin{eqnarray}
D_{\left [ a \right .} Q^{c}_{\,\left . b \right ]} = 0.
\label{Codaccieq}
\end{eqnarray}
\end{definition}

Codacci tensors have received considerable attention
in the mathematics literature (see \cite{CM} for an account)
mainly because of the crucial r\^ole they play in isometric
embeddings of Riemannian manifolds into Euclidean manifolds. In our context
they are interesting because
the full set of constraints for silent universes take the form
of an infinite set of Codacci tensors ${\T{n}}^{a}_{\;b}$ in the
initial data set $(U_p,h_{ab},K_{ab})$ (the symmetry of
${\T{n}}_{ab}$ will be shown later). Let us now consider
the integrability conditions of
the Codacci equation (\ref{Codaccieq}). To that aim, we will 
rewrite this equation as a differential system
for one-forms (this way of writing the
Codacci equation is new, as far as we know).
Let $Q_{ab}$ be a Codacci tensor on  $(\V,\gamma)$.
Choose an arbitrary cobasis $\{\bm{\theta}^a\}$ on this Riemannian manifold
and denote by $\bm{\omega}^a_{\,b}$ the connection one-forms
in this basis. The torsion-free condition is
$d \bm{\theta}^a + \bm{\omega}^{a}_{\;b} \wedge \bm{\theta}^b = 0$.
Denoting by $Q^{a}_{\,b}$ the components of the Codacci tensor $Q$ (with
one index raised) in the basis $\{\bm{\theta}^a \}$ and its dual
$\{\vec{e}_a \}$, we can define three one-forms
$\bm{Q}^a$ by $\bm{Q}^a \equiv Q^a_{\,b} \, \bm{\theta}^b$. A simple 
calculation gives
\begin{eqnarray}
d \bm{Q}^a + \bm{\omega}^{a}_{\,b} \wedge \bm{Q}^b =
\left ( D_{\left [c \right . } Q^{a}_{\, \left . b \right  ]}
\right ) \bm{\theta}^{c} \wedge \bm{\theta}^{b} = 0.
\label{Codform}
\end{eqnarray}
Thus, the Codacci equation for $Q_{ab}$ is equivalent
to $d \bm{Q}^a + \bm{\omega}^{a}_{\,b} \wedge \bm{Q}^b = 0$. 
To study the integrability conditions of this system we only need to
take its exterior derivative, which gives
\begin{eqnarray}
\bm{\Omega}^{a}_{\,\,b} \wedge \bm{Q}^{b} = 0
\label{integ}
\end{eqnarray}
where $\bm{\Omega}^{a}_{\;b}$ are the curvature two-forms
$\bm{\Omega}^{a}_{\;b} = d \bm{\omega}^{a}_{\;b} +
\bm{\omega}^{a}_{\;c} \wedge
\bm{\omega}^{c}_{\;b} $. Let us now assume that $\V$ is three-dimensional
(this is the relevant case for silent universes) and
define the so-called Ricci one-forms by
$\bm{P}_{a} = R_{ab} \,\bm{\theta}^{b}$ ($R_{ab}$ are the components
of the Ricci tensor in the triad $\{ \bm{\theta}^a \}$). The following
identity holds in three dimensions
\begin{eqnarray*}
\bm{\Omega}^{a}_{\,\,b} = \bm{P}^a \wedge \bm{\theta}_b - \bm{P}_{b}
\wedge \bm{\theta}^a - \frac{1}{2} R \, \bm{\theta}^a \wedge \bm{\theta}_b,  
\end{eqnarray*}
where indices are raised and lowered with $h_{ab} = h (\vec{e}_a,
\vec{e}_b )$  (this identity just states the well-known
equivalence between the Riemann and Ricci tensors in three
dimensions). The fact that $Q_{ab}$ is symmetric can be written as
$\bm{Q}^a \wedge \bm{\theta}_{a} = 0$ so that the
integrability conditions (\ref{integ}) become $P_{b} \wedge \bm{Q}^b
 = 0$ or, in index notation,
\begin{eqnarray}
R^{a}_{\;b} \, Q^{b}_{\;c} - Q^{a}_{\;b} \, R^{b}_{\;c} = 0.
\label{commu}
\end{eqnarray}
This equation states that 
$R^{a}_{\;b}$ and $Q^{a}_{\;b}$ commute when viewed as linear maps on the 
tangent space $T_q \V$, $q \in \V$. Let us now put
$(\V,\gamma)=(U_p,h_{ab})$.  Applying (\ref{commu})
to the Codacci tensor $K^{a}_{b}$ we obtain, after using the
definition of $S^{a}_{\;b}$ (\ref{Gausscont}),
\begin{eqnarray*}
S^{a}_{\;b} K^{b}_{\;c} - K^{a}_{\;b} S^{b}_{\;c} = 0.
\end{eqnarray*}
Since all ${\T{n} }^{a}_{\;b}$ are polynomials in $K^{a}_{\;b}$ and  
$S^{a}_{\;b}$, we obtain
${\T{n} }^{a}_{\;b} {\T{m} }^{b}_{\;c} 
- {\T{m} }^{a}_{\;b} {\T{n} }^{b}_{\;c} = 0$, $\forall n,m \in \Bbb{N}
\cup \{0\}$. In particular, each tensor ${\T{n}}^{a}_{b}$ commutes
with $K^{a}_{b}$. The defining recursion 
(\ref{defCod}) then implies that ${\T{n}}_{ab}$ $\forall n \in \Bbb{N}$
are symmetric and hence Codacci tensors.

Obviously, the same arguments hold for an arbitrary value of $t \in I_0$. 
Thus $\T{n}^{a}_{\;b} (t)$ are Codacci tensors
in $(U_p, h_{ab}(t))$ and they commute with each other (for fixed $t$).
Therefore, there exists an orthonormal basis of vectors $\{\hat{e}_a(t) \}$
in $(U_p, h_{ab}(t))$ such that 
$\T{n}^{a}_{\;b}(t)$, $\forall n$ diagonalize simultaneously. 
Denoting by $\{\bm{\hat{\theta}}^a(t) \}$ the dual cobasis, we have 
\begin{eqnarray}
\bm{K}(t) = \sum_{c=1}^{3}
\lambda_0^{c} (t) \, \bm{\hat{\theta}}^c(t) \otimes \hat{e}_{c}(t), 
\hspace{1cm}
\bm{S}(t) = \sum_{c=1}^{3}
\lambda_1^{c} (t) \, \bm{\hat{\theta}}^c(t) \otimes \hat{e}_{c}(t), 
\label{eig}
\end{eqnarray}
where  $\bm{K}(t)$, $\bm{S}(t)$ denote the tensors with
indices $K^a_{\;b}(t)$ and $S^a_{\;b}(t)$, and
$\lambda_0^c(t)$ and $\lambda_1^c(t)$ are their respective
eigenvalues. Writing equation (\ref{dtS}) in this frame  gives
\begin{eqnarray}
\left . \left (\lambda_1^a - \lambda_1^b \right ) \bm{\hat{\theta}}^b \left (
\frac{\partial}{\partial t} \hat{e}_a (t) \right ) \right |_t
 = \left .
\delta^a_{b}
\left [ - \frac{\partial \lambda_1^a }{\partial t} +
3 \lambda_0^a \lambda_1^a - 2 S \lambda_0^a - 2 K \lambda_1^a + SK -
\tr{(SK)} \right ] \right |_t \label{lam02}
\end{eqnarray}
where the Einstein summation convention has been suspended. Until here
no restriction on the Petrov type of the silent universe has
been imposed. However, silent universes
of Petrov type D and 0 are well-understood and attention can be restricted
to Petrov type I. Of course, the
Petrov type of a spacetime need not remain constant everywhere
but it is well-known that the set of points with
Petrov type I is open (see e.g. \cite{Graham}) and therefore a
submanifold, so we can assume Petrov type I everywhere without loss of
generality (because only local properties are considered in this paper).
For silent universes, Petrov type I is equivalent (see e.g.
\cite{BR})  to $E_{ab}(t)$
having three different eigenvalues (of course only two of them are
linearly independent because of the trecefree condition of $E_{ab}(t)$).
Eq. (\ref{Gausscont}) shows that this is also equivalent to
$S_{ab}(t)$ having three different eigenvalues. Thus, 
(\ref{lam02}) for $a \neq b$ gives
\begin{eqnarray}
\bm{\hat{\theta}}^b \left ( \frac{\partial}{\partial t} \hat{e}_a (t)
\right ) = 0, \hspace{1cm} b \neq a, \label{dthet}
\end{eqnarray}
which allows us to rewrite equation (\ref{dtk2}) as
\begin{eqnarray}
\frac{\partial \lambda_0^a (t)  }{\partial t} =
- \left ( \lambda_1^a(t) + \left [\lambda_0^a(t) \right ]^2 \right ). 
\label{lam01}
\end{eqnarray}
We now quote without proof a lemma due to Barnes and Rowlingson \cite{BR}
which we will use in the vacuum case. The proof can be easily rewritten
in our formalism by using the Codacci
equation in the form (\ref{Codform}) 
and the relations (\ref{lam02}),(\ref{dthet}) and (\ref{lam01}).
\begin{lemma}(Barnes and Rowlingson \cite{BR})
Let $(\M, g )$ be a silent universe of Petrov type I and 
$(U_p,h_{ab}(t),K_{ab}(t))$ constructed as in Sect. 2.  Fix an
arbitrary point $q \in U_p$.
Then,
\begin{enumerate}
\item  the subset of $I_0$ where ${\bm K}|_q(t)$ has three different
eigenvalues is dense in $I_0$.
\item Fix an arbitrary value $t \in I_0$.  Each one-form in the
cobasis $\{\bm{\hat{\theta}}^a(t) \}$ introduced above is integrable.
\item There exists a coordinate system $\{x,y,z\}$ in $U_p$ in
which  $\bm{K}(t)$ and $\bm{S}(t)$ are diagonal and such that
the metric $h_{ab}(t)$ takes the form
$ds^2(t) = A (t,x,y,z) dx^2 + 
B (t,x,y,z) dy^2 + C (t,x,y,z) dz^2$.
\end{enumerate}
\label{rank}
\end{lemma}

\section{Vacuum silent universes of Petrov type I}

Vacuum silent universes are characterized by $\rho=0$ which,
from equation (\ref{Gausscont}), is
equivalent to $S(t)=0$. Hence, the evolution equation (\ref{lam02}) 
takes the simpler form
\begin{eqnarray}
\frac{\partial \lambda^a_1(t)}{\partial t} =  \left . \left [
\frac{}{}
3 \lambda^a_1\lambda^a_0 - 2 \lambda^a_{1} K -
\tr{ (S K)} \right ]  \right |_t.
\label{evolvac}
\end{eqnarray}
We need the following lemma
\begin{lemma}
Let $(\M, g )$ be a vacuum silent universe of Petrov type I and 
construct
$(U_p,h_{ab}(t),K_{ab}(t))$ as in Sect. 2.  Fix an
arbitrary point $q \in U_p$. Let $\mu_q(t)$ denote one of the
eigenvalues
$\lambda^a_A |_q (t), A=0,1$ defined in (\ref{eig}). Then, the set
$\{ t\in I_0 ; \mu_q(t) \neq 0\}$ is dense in $I_0$.
\end{lemma}
{\it Proof.} Let first $\mu_q(t)$ be one of the eigenvalues
of $\bm{S}$. We can choose $\mu_q(t)=\lambda^1_1 |_q (t)$ without loss of
generality. Suppose that the lemma does not hold, i.e.
that $\mu_q(t)$ vanishes for $t\in I_1$ where $I_1$ is open and non-empty.
Then, equation (\ref{evolvac}) implies
$\tr{(SK)}|_q (t) = 0$ $\forall t \in I_1$,
which, after using 
$S(t)=0$, implies $(\lambda^2_0(t) - \lambda^3_0(t)) |_q = 0$
$\forall t \in I_1$. This is impossible from the first conclusion
of Lemma \ref{rank}. 
Let now $\mu_q(t)$ be one of the eigenvalues of ${\bm K}(t)$,
say $\lambda^1_0 |_q(t)$. The
claim of the lemma follows easily because if $\mu_q(t)$ vanished on
a non-empty open set $I_1$, then (\ref{lam01}) would imply
$\lambda^1_0 |_q (t)=0$ on $I_1$, which we have just shown to be impossible.
$\hfill \Box$

This lemma combined with Lemma \ref{rank} shows that there
exits an open dense subset $W \subset U_q \times I_0$ where the
eigenvalues $\lambda^a_0(t)$ are non-zero and mutually distinct 
and that the same holds for $\lambda^a_1(t)$.
Then, the following parametrization exists on $W$
\begin{eqnarray*}
\lambda^1_0(t) = w |_t, \hspace{1cm}
\lambda^2_0(t) = w(1+u) |_t,\hspace{1cm}
\lambda^3_0(t) = w(1+v) |_t, \\
\lambda^1_1(t) = w^2 s (k+1) |_t, \hspace{1cm}
\lambda^2_1(t) = w^2 s (k-1) |_t,\hspace{1cm}
\lambda^3_1(t) = -2 w^2 s k |_t, 
\end{eqnarray*}
where $w$, $u$, $v$, $s$ and $k$ are nowhere vanishing scalar functions on
$W$. Furthermore, the combinations
$u -v$, $v+1$, $u+1$, $3k-1$, $3k+1$, $k -1$ and
$k+1$ are also nowhere zero on $W$ (these statements just translate
the fact that $\bm{S}(t)$ and $\bm{K}(t)$ have three different and
non-vanishing eigenvalues on $W$).
This type of parametrization is convenient for the algebra computing
calculations we describe in Appendix A. 
The evolution equations (\ref{lam01}) and (\ref{evolvac}) take
the following form on $W$
\begin{eqnarray}
 \partial_{t} w  =   - w^2 \left (s k + s +1 \right ),  \hspace{1cm}
 \partial_{t} u  =
 w \left (u s k + u s +2 s -u^2 -u \right ),  \nonumber \\
 \partial_{t} v  =    
w \left (\frac{}{} k s v + 3 k s + \left (v+1 \right )\left (s-v \right ) 
 \right ), \hspace{3mm}
\partial_{t} k   =  
 w \left ( - k u + 2 k v - \frac{1}{2} u
+\frac{3}{2} k^2 u \right ), & &  \nonumber \\
 \partial_{t} s   
=   w s \left (2 s k - 1 +2 s - \frac{u}{2} -2 v
- \frac{3}{2} k u \right ) . \label{evolfin}
\end{eqnarray}

We are now in a position to prove our main theorem.
\begin{theorem}
Let $(\M,g)$  be a vacuum silent universe of Petrov type I. Then the 
spacetime is locally isometric to a Kasner spacetime.
\label{main}
\end{theorem}
Before proving this result, let us recall that the 
the Kasner spacetime \cite{Kas} is defined as the manifold 
$\Bbb{R}^{+}\times \Bbb{R}^3$ endowed with the metric 
\begin{eqnarray}
ds^2 = -dt^2 + t^{2p_1} dx^2 + t^{2p_2} dy^2 + t^{2p_3} dz^2,
\hspace{1cm} 
t > 0, \quad-\infty < x,y,z < + \infty
\label{Kasner}
\end{eqnarray}
where $p_1, p_2, p_3 \in \Bbb{R}$ and satisfy the relations
$p_1 + p_2 + p_3 = p_1^2 + p_2^2 + p_3^2 = 1$. Hence each element of
the family is parametrized by one real parameter. This spacetime contains
a three-dimensional abelian isometry group acting transitively on the 
spacelike hypersurfaces $t = \mbox{const.}$ Is is thus a Bianchi type
I vacuum spacetime. Furthermore, it is well-known (see e.g. \cite{KSMH}) that
any spacetime admitting an abelian three-dimensional Lie algebra of
Killing vectors which span spacelike hypersurfaces must be locally isometric
to Kasner (a global isometry requires further conditions on the
topology of the spacetime). Hence, the Kasner family can be locally
characterized by the existence of these Killing vectors. Theorem \ref{main}
provides another characterization for the Kasner family directly in terms
of the Weyl tensor and without
involving isometries, namely, {\it a vacuum spacetime which is of Petrov type I
and silent must be locally isometric to Kasner}.
Obviously a global isometry cannot
be expected because the Petrov type I and the silent conditions are purely
local and they place no restriction on the global topology of the spacetime 
(which could be, for instance, $\Bbb{R} \times T^3$, where $T^3$ is the
three-dimensional torus, endowed with the metric (\ref{Kasner})).

The proof of Theorem \ref{main} involves combining several constraint
equations in order to show that they lead to incompatibilities
unless the spacetime is very simple, namely a homogeneous Bianchi
model. Since the expressions involved soon become very large, the
proof would be impossible (using this direct method) without employing
algebraic computing. However, the proof can be followed
in exactly the same way as an ordinary proof can. The only
requirement is some knowledge of Reduce, which is the algebraic 
computing program we have used.
Even for those who are not familiar with Reduce, we believe that the
general idea of the proof can be followed with relatively little effort.
Hence, the Reduce program and the necessary explanations 
on the logic of the proof have been included in Appendix A
(the meaning of the commands is not explained, for the interested
reader we recommend the introduction to Reduce by M.A.H.MacCallum and F.Wright
\cite{Mac}).

{\it Proof of the Theorem.}

Let us consider a
point $p\in \M$ and construct
$(U_p,h_{ab}(t),K_{ab}(t))$, $t\in I_0$ as described in Sect 2.
Using Lemma \ref{rank}, the metric $h_{ab}(t)$
in $U_p$ can be written as
\begin{eqnarray}
ds^2(t)  = A(t,x,y,z) dx^2 + B(t,x,y,z) dy^2 + C(t,x,y,z) dz^2.
\label{onepa}
\end{eqnarray}
In Appendix A we proof that, under the assumptions of the theorem, the
functions $A$, $B$ and $C$ 
depend only on $t$. Then, the spacetime metric in $(\M_p, g|_{\M_p})$
can be reconstructed from $h_{ab}(t)$ in the usual way
to give $ds^2 |_{\M_p} = -dt^2 + A(t) dx^2 + B(t) dy^2 + C(t) dz^2$. 
Thus, the spacetime admits three commuting Killing vectors which
span spacelike hypersurfaces. Hence, the metric must 
must take the form (\ref{Kasner}) for some values of $p_1$, $p_2$ and
$p_3$. Conversely, a simple calculation shows that
the Kasner spacetime is a silent universe of Petrov type I. This 
concludes the proof of the theorem. $\hfill \Box$

\vspace{5mm}

{\bf Acknowledgement}

\vspace{5mm}

I am very grateful to J.M.M Senovilla for useful
discussions and to R.Beig for his interest and suggestions.
I also wish to thank the European Union
for financial support under the Marie Curie fellowship ERBFMBICT972520.

\section*{Appendix}

This Appendix contains the bulk of the proof of Theorem {\ref{main}.
The aim is to prove that the one-parametric
family of metrics (\ref{onepa}) on $U_p$ (for vacuum silent
spacetimes of Petrov type I) can be written in the form
$ds^2 (t) = A(t) dx^2 + B(t) dy^2 + C(t) dz^2$.
In order to prove this, we use the algebraic computing program
Reduce. In this Appendix we include the code lines, which
are written using the {\sf sans serif} font, and we explain 
the logic of the proof (this is written in plain text). 
A few key output results\footnote{In Reduce,
a code line ending with \$ does not
produce any written output (of course the operation is performed).
If the line ends with ``;'' the result is written in the screen (or the
standard output).}, which are required to follow the proof,
are also included. These will be written in {\it italic} characters.

\vspace{2mm} 

{\sf load-package groebner\$ on gcd\$

depend w,t,z\$ depend u,t,z\$ depend v,t,z\$ depend s,t,z\$
depend k,t,z\$ }

\vspace{2mm}

Only the dependence on $t$ and $z$ of the functions is prescribed.
Of course, all of these functions depend on $y$ and $x$ as well,
but we will not invoke any equations containing partial
derivatives with respect to $x$ or $y$. This may seem
inadequate because we are not exploiting all the available
equations. However, the strategy of the proof relies on showing a certain
property for the $z$ variable (the vanishing of some associated Christoffel
symbols) and then use the property that the $z$ variable is on the
same footing as the $x$ and $y$ variables (there is an obvious symmetry
between $x <-> y <-> z$ in the metric (\ref{onepa})) in order to imply the
vanishing of some other Christoffel symbols. 
Hence concentrating only on the $z$ variable is sufficient.

Let us now introduce the evolution equations (\ref{evolfin}) and the set
of Codacci tensors $\T{n}$.

\vspace{2mm}

{\sf 
s(0,1):=-w\$ s(0,2):=-w$\bm{\ast}$(1+u)\$ s(0,3):=-w$\bm{\ast}$(1+v)\$
s(1,1):=w\^{}2$\bm{\ast}$s$\bm{\ast}$(1+k)\$ 

s(1,2):=w\^{}2$\bm{\ast}$s$\bm{\ast}$(k-1)\$ s(1,3):=-2$\bm{
\ast}$w\^{}2$\bm{\ast}$s$\bm{\ast}$k\$ 

dwt:=-w\^{}2$\bm{\ast}$(s$\n$k+s+1)\$
dut:= w$\bm{\ast}$(u$\bm{\ast}$s$\bm{\ast}$k+u$\bm{\ast}$s+2$\bm{
\ast}$s-u\^{}2-u)\$

dvt:= w$\bm{\ast}$(v$\bm{\ast}$s$\bm{\ast}$k+v$\bm{\ast}$s+s+3$\bm{
\ast}$k$\bm{\ast}$s-v\^{}2-v)\$

dst:= w$\bm{\ast}$s$\bm{\ast}$(2$\bm{\ast}$s$\bm{\ast}$k-1+2$\bm{
\ast}$s-u/2-2$\bm{\ast}$v-3/2$\bm{\ast}$k$\bm{\ast}$u)\$

dkt:= w$\bm{\ast}$(-k$\bm{\ast}$u+2$\bm{\ast}$k$\bm{\ast}$v-u/2+3/2$\bm{
\ast}$k\^{}2$\bm{\ast}$u)\$

df(w,t):=dwt\$ df(u,t):=dut\$ df(v,t):=dvt\$ df(s,t):=dst\$ df(k,t):=dkt\$ 

d:=6\$ for l:=2:d do for i:=1:3 do s(l,i):=df(s(l-1,i),t)-s(l-1,i)$\bm
{\ast}$s(0,i)\$ }

\vspace{2mm}

The symbol $s(l,i)$, where $l \in \Bbb{N} \cup \{0\}$ 
and $i=1,2,3$, is used to denote the $i-th$ diagonal coefficient
of the Codacci tensor $\T{l}$. Let us introduce the Codacci equations
(we denote the Christoffel symbols $\Gamma^{i}_{jk}$ of
the metric (\ref{onepa}) by gam(i,j,k), $i,j,k=1,2,3$).

\vspace{2mm}

{\sf
for l:=0:d do $\bm{<<}$ cdz(l,1):= df(s(l,1),z)+gam(1,1,3)$\bm{
\ast}$(s(l,1)-s(l,3))\$

\hspace{26mm} cdz(l,2):= df(s(l,2),z)+gam(2,2,3)$\bm{
\ast}$(s(l,2)-s(l,3)) $\bm{>>}$ \$

df(w,z):=rhs first solve(cdz(0,1),df(w,z))\$
df(u,z):=rhs first solve(cdz(0,2),df(u,z))\$

df(k,z):=rhs first solve(cdz(1,1),df(k,z))\$
df(s,z):=rhs first solve(cdz(1,2),df(s,z))\$

df(v,z):=rhs first solve(cdz(2,1),df(v,z))\$

for i:=0:(d-3) do r(i):=num cdz(i+3,2)/w\^{}(4+i)/s\$

for i:=0:(d-4) do n(i):=resultant(r(0),r(i+1),gam(1,1,3))/2 \$ }

\vspace{2mm}

Our next aim is to analyze the case in which gam(2,2,3)$\equiv 0$
on a non-empty open subset $\tilde{W} \subset W$ and gam(1,1,3) is
nowhere zero on $\tilde{W}$.

\vspace{2mm}

{\sf gam(2,2,3):=0\$

pro:=r(0)/gam(1,1,3)\$ proz:=df(pro,z)/gam(1,1,3)\$
prozz:=df(proz,z)/gam(1,1,3)\$

n1:=-2$\bm{\ast}$v\^{}2+12$\bm{\ast}$v$\bm{\ast}$(k-1)-9$\bm{
\ast}$k\^{}2+30$\bm{\ast}$k-13\$ n2:=9$\bm{\ast}$k-4$\bm{\ast}$v-7\$

factorize\{prozz-n1$\bm{\ast}$pro-n2$\bm{\ast}$proz\}; }

${\it \{18, -v, v, 3k + 1, 3k + 1, k - 1\} }$

\vspace{2mm}

But this is impossible because no element in this list
can vanish anywhere on $W$. This concludes this case. 
Next, we consider the situation in which
gam(2,2,3) is non-zero almost everywhere on $W$ (studying this case
constitutes the heart of the proof).
Possibly after restricting $W$ to
a smaller dense open subset, we can assume  
gam(2,2,3) $\neq 0$ everywhere on $W$ (from now on we will 
restrict $W$ to an open dense subset thereof whenever necessary and without
further notice).

\vspace{2mm}

{ \sf 
clear gam(2,2,3)\$ pol1:=-n(0)/gam(2,2,3)\$ pol2:=n(1)/gam(2,2,3)+5$\bm{
\ast}$u$\bm{\ast}$pol1\$

l1:=22$\bm{\ast}$(  585$\bm{\ast}$k\^{}2$\bm{\ast}$s - 192$\bm{\ast}$k$\bm{
\ast}$s - 25$\bm{\ast}$s+360$\bm{\ast}$k$\bm{\ast}$u)\$

l2:=-1989$\bm{\ast}$k\^{}2$\bm{\ast}$u - 2712$\bm{\ast}$k$\bm{
\ast}$u - 7008$\bm{\ast}$k - 275$\bm{\ast}$u\$

pol3:=-264$\bm{\ast}$k$\bm{\ast}$n(2)/gam(2,2,3)+l1$\bm{\ast}$pol1+l2$\bm{
\ast}$pol2\$}

\vspace{2mm}

A number of particular cases must be analyzed.
Let us start by studying what happens 
when the polynomial $c_0$, defined immediately below, vanishes on a
non-empty open subset $W_1 \subset W$.

{\bf Subcase 0:} $c_0=0$ identically on $W_1$.
\vspace{2mm}

{\sf c0:= - 27$\bm{\ast}$k\^{}4$\bm{\ast}$u\^{}2 - 72$\bm{
\ast}$k\^{}3$\bm{\ast}$v$\bm{\ast}$u + 36$\bm{\ast}$k\^{}3$\bm{
\ast}$u\^{}2 + 96$\bm{\ast}$k\^{}2$\bm{\ast}$v\^{}2 - 96$\bm{
\ast}$k\^{}2$\bm{\ast}$v$\bm{\ast}$u

+42$\bm{\ast}$k\^{}2$\bm{\ast}$u\^{}2 - 24$\bm{\ast}$k$\bm{\ast}$v$\bm
{\ast}$u + 12$\bm{\ast}$k$\bm{\ast}$u\^{}2 + u\^{}2\$

c0t:=df(c0,t)/2/w\$ c01:=resultant(c0t,pol1,s)\$

rg:=resultant(c0,c01,v)/(3$\bm{\ast}$k+1)\^{}12/(3$\bm{
\ast}$k-1)\^{}12/k\^{}12/3131031158784/u\^{}14\$

rg+(-117$\bm{\ast}$k\^{}4+30$\bm{\ast}$k\^{}2-1 )\^{}2; }

${\it { 0 }}$ 

\vspace{2mm}

So, $\left . -117k^4+30k^2-1 \right |_{W_1} \equiv 0$. In
particular $k$ is constant and its derivative is identically zero.
Then

\vspace{2mm}

{\sf part(groebner(\{dkt/w$\bm{
\ast}$2,-117$\bm{\ast}$k\^{}4+30$\bm{\ast}$k\^{}2-1,c0\}),4); }

${\it  u^2} $ 

\vspace{2mm}

which is impossible because $u$ cannot vanish anywhere on $W$.
So we can assume $c_0 \neq0$ everywhere on $W$ and divide out this factor
whenever necessary. This finishes Subcase 0. We continue with the generic
case.

\vspace{2mm}

{\sf v:=u$\bm{\ast}$l\$ u1:=resultant(pol1,pol2,s)/(3$\bm{
\ast}$k+1)/(3$\bm{\ast}$k-1)/(u-v)/v/u\^{}7/c0/324\$

u2:=resultant(pol1,pol3,s)/c0/162/v/u\^{}6/(u-v)\$}

\vspace{2mm}

By construction, the polynomials $u_1$ and $u_2$ depend only on
$k,l$ and $u$.

\vspace{2mm}

{\sf q1:=sub(k=2,l=2,u1)\$ length coeff(q1,u); length coeff(u1,u);}

${\it   { 5 } \hspace{3mm}  { 5 }}$
    
{\sf q2:=sub(k=2,l=2,u2)\$ length coeff(q2,u); length coeff(u2,u); }

${\it  {7} \hspace{5mm}  {7}}$

\vspace{2mm}

The two polynomials $q_1$ and $q_2$ depend only on $u$. The two
last output results show that $q_1$ and $u_1$ are of the same degree as
polynomials in $u$ and the same happens for $q_2$ and $u_2$.

\vspace{2mm}

{\sf groebner\{q1,q2\};}

${\it  \{1\} }$

\vspace{2mm}

Hence $q_1$ and $q_2$ have no common zeros. From this
we can conclude that the resultant of $u_1$ and $u_2$ with
respect to $u$ is not identically zero. Indeed, if
it were zero, this would mean that the polynomials
$u_1$ and $u_2$ have common solutions for $u$ at any
value of $k$ and $l$. In principle, it could happen that
the common solution for $u$ tends to $\infty$ when we approach
$k=2$ and $l=2$, but this can also be excluded because
$q_1$ and $u_1$ are polynomials in $u$ of the same degree (and 
similarly for $q_2$ and $u_2$). So, there should exist at least
one finite value of $u$ which solves $q_1=0$ and $q_2=0$ simultaneously.
But we have proven that $q_1$ and $q_2$ have no common zeros. Thus, 
the resultant of $u_1$ and $u_2$ with
respect to $u$ is a non-identically vanishing polynomial in $l$ and $k$.
The reason why we do not prove this by direct calculation is
that the polynomials are already very big and performing 
the full resultant takes far too long.

So, $k$ and $l$ must belong 
to the set of zeros of a polynomial in $k$ and $l$, which we 
denote by $\sigma$
(we view $\sigma$ as
as a subset of the $\{k,l\}$ plane). Next, we prove that
restricting the polynomial $u_1$ to any point on $\sigma$ provides
a non-trivial polynomial in $u$ (i.e. non-constant).
Similarly, we show that  $pol_2$ (which depends on $k,l,u,s$)
when restricted to any point on $\sigma$ is a non-trivial polynomial in $s$.
These two facts imply that $\{k,l,u,s\}$ must lie on a one-dimensional
manifold. So, let us first prove that $u_1$ is a non-trivial
polynomial in $u$ at any point on $\sigma$.

\vspace{2mm}

{\sf  j1:=lterm(u1,u)/u\^{}4\$ j2:=lterm(u2,u)/u\^{}6\$

j3:=sub(u=0,u1)/(3$\bm{\ast}$k+1)/(3$\bm{\ast}$k-1)\$

j4:=sub(u=0,u2)/(3$\bm{\ast}$k+1)/(3$\bm{\ast}$k-1)/414720$\bm{
\ast}$u\^{}2/k\^{}2/c0\$

g:=groebner\{j3,j4\}\$ g2:=second(g)/l\^{}2/(l-1)\^{}2\$

y1:=resultant(j1,g2,l)/(k-1)\^{}3/(k+1)\^{}5/k\^{}12/(3$\bm{
\ast}$k+1)\^{}7/(3$\bm{\ast}$k-1)\^{}7/471859200\$

y2:=resultant(j2,g2,l)/3774873600/(3$\bm{\ast}$k-1)\^{}6/(3$\bm{
\ast}$k+1)\^{}6/(k+1)\^{}5/(k-1)\^{}3/k\^{}14\$

groebner\{y1,y2\}; }

$ {\it \{1\}}$

\vspace{2mm}

which proves that $u_1$ is nowhere identically zero on $\sigma$. Since
a constant non-zero value is also impossible (recall that $u_1=0$ is
a consequence of the Codacci equations) the claim follows. Regarding
$pol_2$, it suffices to notice that the leading term (in the variable
$s$) of this polynomial is

\vspace{2mm}

{\sf lterm(pol2,s)\$ on factor\$ ws; off factor\$}

$ {\it -44 (3 k+1)^2 (3 k-1)^2 s^3}$

\vspace{2mm}

which is nowhere zero on $W$.
Hence, $u, k, v, s$ lie on a curve. Therefore, there
exists a non-zero
vector field $\vec{a}=a_1\partial_z + a_2  \partial_t$
($a_1$ and $a_2$ unknown functions on $W$) which annihilates $k, s,
u, v$. Consider first the possibility that $a_1$ vanishes
on a non-empty open set, i.e. suppose that $k,s,u,v$ do not depend on $t$ on
that open set. Then

\vspace{2mm}

{\sf v:=rhs first solve(2$\bm{\ast}$dkt/w,v)\$
s:=rhs first solve(2$\bm{\ast}$k$\bm{\ast}$dst/w/s,s)\$

third groebner({num dut/w,num dvt/w,num pol1/u,num pol2},{u,k})/(3$\bm{
\star}$k+1)\^{}2/

(3$\bm{\star}$k-1)\^{}2/(k+1)\^{}3/k; }

${\it 4k^2-1}$ \hspace{5mm} Hence $k$ is constant.

{\sf first groebner(\{num dut/w,num dvt/w,4$\bm{\ast}$k\^{}2-1\},\{u,k\});}

${\it \{8 k + 3u +8,4k^2-1\}}$ \hspace{5mm} So, $u$ is also constant. 
In particular, its derivative with respect to $z$ vanishes.

{\sf first groebner\{df(u,z),r(0),2$\bm{\ast}$u+8+8$\bm{
\ast}$k,4$\bm{\ast}$k\^{}2-1\}; clear s\$ clear v\$ }

${\it \mbox{gam}(2,2,3)}$

This is impossible in the case we are analyzing.
So, we can assume without loss of generality that
the vector $\vec{a}$ takes the form
$\vec{a}= \partial_z - b \cdot \mbox{gam}(2,2,3)/w \partial_t$
for some unknown function $b$. This provides a new set of equations
(we also rewrite gam(1,1,3) $=f \bm{\ast}$gam(2,2,3) where
$f$ is an unknown function)

\vspace{2mm}

{\sf 
clear v\$ gam(1,1,3):=gam(2,2,3)$\bm{\ast}$f\$

pr(1):=2$\bm{\ast}$(df(k,z)-b$\bm{\ast}$gam(2,2,3)$\bm{
\ast}$df(k,t)/w)/gam(2,2,3)\$

pr(2):=2$\bm{\ast}$(df(s,z)-b$\bm{\ast}$gam(2,2,3)$\bm{
\ast}$df(s,t)/w)/gam(2,2,3)/s\$

pr(3):=(df(u,z)-b$\bm{\ast}$gam(2,2,3)$\bm{\ast}$df(u,t)/w)/gam(2,2,3)\$

pr(4):=(df(v,z)-b$\bm{\ast}$gam(2,2,3)$\bm{\ast}$df(v,t)/w)/gam(2,2,3)\$

pr(5):=r(0)/gam(2,2,3)\$

for i:=1:4 do x(i):=resultant(pr(1),pr(i+1),f)\$

z1:=-resultant(x(1),x(2),b)/(k-1)/(3$\bm{\ast}$k+1)/2\$

z2:=(-2$\bm{\ast}$resultant(x(1),x(3),b)/(k-1)/(3$\bm{\ast}$k+1)/2+(3$\bm{
\ast}$k+1)$\bm{\ast}$z1)/(3$\bm{\ast}$k-1)\$

z3:=-resultant(x(1),x(4),b)/(k-1)/(3$\bm{\ast}$k+1)/2\$

f1:=  ( 9$\bm{\ast}$k\^{}2$\bm{\ast}$u - 24$\bm{\ast}$k$\bm{\ast}$u + 30$\bm
{\ast}$k$\bm{\ast}$v - 9$\bm{\ast}$u + 6$\bm{\ast}$v)/2\$

f2:= ( -9$\bm{\ast}$k\^{}2$\bm{\ast}$u - 6$\bm{\ast}$k$\bm{
\ast}$u - 36$\bm{\ast}$k$\bm{\ast}$v - 36$\bm{\ast}$k + 3$\bm{
\ast}$u + 12$\bm{\ast}$v + 12)/2\$

z3:=(z3-f1$\bm{\ast}$z1-f2$\bm{\ast}$z2)/(3$\bm{\ast}$k-1)\$ }

\vspace{2mm}

Before following with the general case, we must study a few particular
cases. They are characterized by the vanishing of one of the following
expressions on a non-empty open subset $W_2 \subset W$

\vspace{2mm}

{\sf  c1:= 2$\bm{\ast}$k$\bm{\ast}$v + 3$\bm{\ast}$k + 2$\bm{\ast}$v + 1\$

c2:= 2$\bm{\ast}$k$\bm{\ast}$v+3$\bm{\ast}$k+1\$

c3:= - 3$\bm{\ast}$k\^{}2  + 2$\bm{\ast}$k$\bm{\ast}$v + 2$\bm{
\ast}$k + 2$\bm{\ast}$v + 1\$

c4:=- 4$\bm{\ast}$k\^{}3$\bm{\ast}$v + 3$\bm{\ast}$k\^{}3  - 10$\bm{
\ast}$k\^{}2$\bm{\ast}$v - 17$\bm{\ast}$k\^{}2 - 4$\bm{\ast}$k$\bm{
\ast}$v - 3$\bm{\ast}$k + 2$\bm{\ast}$v + 1\$}

\vspace{2mm}

{\bf Subcase 1:}  $c_1$ vanishes identically on $W_2$.

\vspace{2mm}

{\sf c1t:=2$\bm{\ast}$df(c1,t)/w\$ v:=rhs first solve(c1,v)\$ c1t:=c1t/(3$\bm{
\ast}$k+1)$\bm{\ast}$(k+1)\^{}2\$

u:=rhs first solve(c1t,u)\$ r1:=-z2$\bm{\ast}$8$\bm{
\ast}$(k-1)\^{}2/(k+1)/(3$\bm{\ast}$k+1)\$

r1-(2$\bm{\ast}$k$\bm{\ast}$s+2$\bm{\ast}$s-1)\^{}2$\bm{\ast}$(5$\bm{
\ast}$k-1)\^{}2; }\$

 ${\it 0}$ \hspace{5mm}  So, either $k=1/5$ or 
$2ks +2s -1=0$ on some non-empty open set. $k=1/5$ is impossible
as the following calculation shows

\vspace{2mm}

{\sf  k:=1/5\$ groebner\{dkt/w,pol1\}; clear k\$ } 

 ${\it \{1\}}$   \hspace{5mm} Thus,

{\sf  s:=1/2/(k+1)\$ factorize first groebner\{pol1,pol2\}; 
clear v\$ clear s\$ clear u\$ }

 $ {\it  \{3k + 1,3k+1 \} }$

\vspace{2mm}

This is also impossible and therefore $c_1$ cannot vanish on $W_2$.
As usual, we can assume that it is nowhere zero on $W$.

{\bf Subcase 2:} $c_2$ vanishes identically on $W_2$.

\vspace{2mm}

{\sf 
c2t:=df(c2,t)/w$\bm{\ast}$2\$ gr:=groebner(\{z1,c2,c2t\},\{s,u,v\})\$

s:=rhs first solve(first gr,s)\$
u:=rhs first solve(second gr,u)\$

v:=rhs first solve(third gr,v)\$

length coeff(num pol1,k); clear u\$ clear v\$ clear s\$ }

 ${\it \{8\} }$

\vspace{2mm}

Thus, $k$ must by constant and hence $s$, $v$ and $u$ are
also constants. In particular they do not depend on $t$ and we have seen
above that this is impossible for gam(2,2,3)$\neq 0$.

{\bf Subcase 3:} $c_3$ vanishes identically on $W_2$.

\vspace{2mm}

{\sf c3t:=df(c3,t)/w\$ gr:=groebner(\{c3,c3t\},\{s,v\})\$

v:=rhs first solve(second gr,v)\$
s:=rhs first solve(first gr,s)\$

cc:=factorize (num z1/(3$\bm{\ast}$k+1))\$ 

length cc;}

${\it 2}$

{\sf 
second groebner(\{first cc,z2\},\{u,k\})\$ on factor\$ ws; off factor\$}

 ${\it (3 k^2  + 1)(k^2  + 1)(3k + 1)(3k - 1)(k - 1)}$

{\sf 
second groebner(\{second cc,pol1,z2\},\{u,k\})\$ on factor\$ ws; off factor\$}

 $ {\it (3k + 1)(3k - 1)^2(k - 1)} $

{\sf clear u\$ clear v\$ clear s\$ }

\vspace{2mm}

This gives a contradiction to $c_3=0$ on $W_2$, hence  $c_3$ is
non-zero everywhere on $W$.

{\bf Subcase 4:} $c_4$ vanishes identically on $W_2$

\vspace{2mm}

{\sf c4t:=2$\bm{\ast}$df(c4,t)/w\$ groebner(coeff(c4,v)); }

 $ { \it \{1\}} $ \hspace{5mm}  This shows that we can solve for $v$

{\sf  v:=rhs first solve(c4,v)\$

groebner(\{z1,z2,5$\bm{\ast}$k\^{}2-1,c4t\}); }

 $ {\it \{1\}} $ \hspace{5mm} We can solve $s$ in c4t

{\sf 
s:=rhs first solve(c4t,s)\$ length coeff(resultant(num z1,num z2,u),k); }

 $  {\it 52} $  \hspace{5mm}  Hence, $k$ must be constant and its
derivative must be identically zero. Then

{\sf 
second groebner(\{dkt,z1,z2\},\{u,k\})/num(v)\$ on factor\$ ws;  off factor\$ 

clear v\$ clear s\$ }

 $ {\it (k^2  + 1)(3k - 1)}$ 

\vspace{2mm}

Again a contradiction. This concludes the last particular case. 

{\bf Generic Case:} We can now
consider the generic situation (we divide by $c_1$, $c_2$, $c_3$ and $c_4$
whenever necessary)

\vspace{2mm}

{\sf 
h(1):=resultant(z1,z2,s)/c1/2\$ h(2):=resultant(z1,z3,s)/c1/2\$

h(3):=resultant(z1,pol1,s)\$

t(1):=resultant(h(1),h(2),u)/9216/k\^{}4/(k+1)\^{}2/(3$\bm{
\ast}$k-1)\^{}2/(3$\bm{\ast}$k+1)\^{}2/c2/c3\^{}2/

c4\^{}2/(k-1)/(v+1)/v\$

t(2):=resultant(h(1),h(3),u)/442368/k\^{}6/(3$\bm{
\ast}$k-1)\^{}4/v\^{}3/(3$\bm{\ast}$k+1)\^{}3/c3\^{}2/c4\^{}2/

(v+1)/(k+1)\$}

\vspace{2mm}

As happened above, evaluating the resultant of $t(1)$ and
$t(2)$ with respect to $u$ is unworkable 
and we must use the same type of argument employed above in order to
show that this resultant is not identically zero.
Consider the particular value $k=2$ in
$t(1)$ and $t(2)$. The resulting polynomials, called $tt(1)$
and $tt(2)$ are shown to be of the
same degree as $t(1)$ and $t(2)$ respectively (as polynomials in $u$). 
We then show that $tt(1)$ and $tt(2)$ have 
no common zeros. Hence the resultant of $t(1)$ and $t(2)$ with respect
to $u$ is not identically zero.

\vspace{2mm}
 
{\sf tt(1):=sub(k=2,t(1))\$ length coeff(tt(1),v);}  $  {\it 9}$  
{\sf length coeff(t(1),v);}  $  {\it 9}$ 

{\sf 
tt(2):=sub(k=2,t(2))\$ length coeff(tt(2),v);}  $  {\it 14}$
{\sf  length coeff(t(2),v);}  $ {\it 14}$
 
{\sf groebner\{tt(1),tt(2)\}; }

 $ {\it \{1\} }$

\vspace{2mm}

Since the resultant(t(1),t(2),u) depends
only on $k$ and it is non-zero, we can conclude that $k$ must be constant.
It only remains to show that $v$, $u$ and $s$ are also constants.
This is shown in the next few lines

\vspace{2mm}

{\sf  groebner(coeff(t(1),v)); }

 $  {\it \{1\} }$  \hspace{5mm} So, $v$ must be constant

{\sf factorize lterm(h(1),u); }

 $  {\it \{6,u,u,u,k,k - 1,k + 1,3k + 1\} }$
\hspace{5mm}   $u$ must also be constant

{\sf factorize lterm(z3,s); }

 $  {\it \{32,- s,s,k,k + 1,3k + 1\}}$   \hspace{5mm}
$s$ is also constant.

\vspace{2mm}

But $k, s, u, v$ constants is impossible when
gam(2,2,3)$\neq 0$ (because, in particular, their time derivative
should vanish and we have excluded this case before).
Thus, it follows that gam(2,2,3) $\neq 0$ on $W$ is impossible.
At the very beginning we also showed
that gam(2,2,3) $=0 $ and gam(1,1,3) $\neq 0$ on a non-empty open
subset is impossible. 
Hence, gam(1,1,3) $=$ gam(2,2,3) $=0$ everywhere on $W$.
Furthermore, there is
an obvious symmetry between the coordinates $x, y$ and $z$ (there
is nothing special about $z$). It follows necessarily
gam(1,1,3) $=$gam(2,2,3) $=$  gam(1,1,2) $=$gam(3,3,2) $=$  
gam(2,2,1) $=$gam(3,3,1) $= 0$ everywhere on $W$. Since $W$
is dense on $U_p \times I_0$ (we have restricted $W$ always to smaller
open {\it dense} subsets thereof) all these Christoffel symbols
vanish identically on $U_p$ $\forall t \in I_0$. In terms
of the metric (\ref{onepa}) at $t=0$, this means
$ds^2 |_{t=0} = A(x) dx^2 + B(y) dy^2 + C(z) dz^2$.
Performing a trivial coordinate change we obtain
$ds^2 |_{t=0} = d\tilde x^2 + d\tilde y^2 + d\tilde z^2$.
The Codacci equations 
$D_{\left [ a \right .} K^{c}_{\,\left . b \right ]} = 
D_{\left [ a \right .} S^{c}_{\,\left . b \right ]} = 0$ imply that
$K^{a}_{b}$, $S^a_{b}$ are diagonal constant matrices
in the coordinate system  $\{ \tilde{x},\tilde{y},\tilde{z} \}$.
Then, the evolution equations (\ref{dtk2}), (\ref{dtS}) imply that
$K^a_{b}(t)$ and $S^a_{b}(t)$ are independent of $\tilde{x},\tilde{y}$,
$\tilde{z}$. Finally, the evolution equation (\ref{dth}) shows that $h_{ab}(t)$
is diagonal and depends only on $t$ (in the coordinate system $\{\tilde{x},
\tilde{y},\tilde{z} \}$). After dropping the
tildes we readily obtain $ds^2 (t) = A(t) dx^2 + B(t) dy^2 + C(t) dz^2$,
which was the aim of this Appendix.

\end{document}